\documentclass[%
 a4paper,reprint,onecolumn,
 notitlepage,showkeys,
superscriptaddress,nofootinbib,
 amsmath,amssymb,
 aps,
prd,
floatfix,
]{revtex4-2}
\usepackage{epsfig}
\usepackage[x11names]{xcolor}
\usepackage{hyperref}
\usepackage{orcidlink}
\hypersetup{           
    colorlinks=true,                
    breaklinks=true,                
    urlcolor= Purple2,                
    linkcolor= red,                
    bookmarksopen=false,
    filecolor=black,
    citecolor=SpringGreen4,
    linkbordercolor=blue}
\usepackage{graphicx}
\usepackage{orcidlink}
\usepackage[T1]{fontenc} 
\usepackage[utf8]{inputenc} 

\begin{document}
\title{Thermodynamics and Null Geodesics of a Schwarzschild Black Hole Surrounded by a Dehnen Type Dark Matter Halo}

\author{Mrinnoy M. Gohain\orcidlink{0000-0002-1097-2124}}
\email{mrinmoygohain19@gmail.com}
\affiliation{%
 Department of Physics, Dibrugarh University, Dibrugarh \\
 Assam, India, 786004}

\author{Prabwal Phukon \orcidlink{0000-0002-4465-7974}}%
 \email{prabwal@gmail.com}
\affiliation{%
 Department of Physics, Dibrugarh University, Dibrugarh \\
 Assam, India, 786004}%
 \affiliation{Theoretical Physics Divison, Centre for Atmospheric Studies, Dibrugarh University, Dibrugarh, Assam, India 786004}
 
\author{Kalyan Bhuyan\orcidlink{0000-0002-8896-7691}}%
 \email{kalyanbhuyan@dibru.ac.in}
\affiliation{%
 Department of Physics, Dibrugarh University, Dibrugarh \\
 Assam, India, 786004}%
 \affiliation{Theoretical Physics Divison, Centre for Atmospheric Studies, Dibrugarh University, Dibrugarh, Assam, India 786004}

\begin{abstract}
In this work, we derive a novel black hole solution surrounded by a Dehnen-(1,4,0) type dark matter halo by embedding a Schwarzschild black hole within the halo, constituting a composite dark matter-black hole system. The thermodynamics of the resulting effective black hole spacetime is then studied with particular attention to the influence of the dark matter parameters on various thermodynamic properties. We examine the specific heat and free energy to assess the thermodynamic stability of the model. Furthermore, the null geodesics and the effective potential of light rays are studied to investigate how the dark matter parameters affect these geodesics and the radii of circular orbits. The stability of circular null geodesics is analysed using dynamical systems and Lyapunov exponents, which represent the dynamical behaviour of the circular photon orbits. Finally, we studied if the circular geodesics exhibit chaotic behaviour using the chaos-bound condition.
\end{abstract}

\maketitle

\section{Introduction}
\label{intro}
Black holes (BHs) are no doubt one of the most intriguing theoretical and astrophysical objects currently being studied from various perspectives. One such perspective is black hole thermodynamics (BHT) \cite{Page2005Sep,Bardeen1973Jun}. The thermodynamic features of BHs qualify as a fascinating area for study in BH physics. Thermal fluctuations, often known as Hawking radiation \cite{Hawking1974Mar,Hawking1971May,Page2005Sep,Wald2001Jul,Bardeen1973Jun, Hawking1975Aug}, are interesting speculated phenomena related to BHs. According to Hawking's theory of BH radiation, BHs are not entirely black but rather emit radiation at a temperature inversely proportional to mass. This radiation is caused by quantum mechanical processes that occur at the event horizon, where virtual particles are constantly created and annihilated \cite{Hawking1975Aug}. Due to gravitational effects, these particles can occasionally exit the BH as radiation, resulting in the BH's progressive loss of mass. Thus, BHT plays an important role in providing positive hints about a quantum theory of gravity. The laws of BH mechanics suggest that BHs can be treated as thermodynamic systems similar to the usual laws of thermodynamics. The thermodynamics of BHs also seem to have analogous aspects like temperature, entropy, free energy, etc. Some of the recent works in this field have been carried out by different authors in various frameworks. For example, Ditta et al. \cite{Ditta2023Jun} investigated the thermodynamics of charged torus-like BHs. Yasir et al. \cite{Yasir2024Jan} studied acoustic Schwarzschild BHs in extended thermodynamic phase space admitting phase transition. Mahapatra and Banerjee \cite{Mahapatra2023Feb} studied the thermodynamics of rotating hairy BHs obtained through gravitational coupling. Hendi et al. \cite{Hendi2021Dec} investigated the thermodynamics and phase transitions of a four-dimensional rotating Kaluza-Klein BH in the framework of Maxwell electrodynamics. Singh et al. \cite{Singh2022Mar} derived a BH solution in Lee-Wick gravity having a point source in a higher-derivative theory and then analysed the thermodynamics of such a BH system. Simovic and Soranidis \cite{Simovic2024Feb} studied the thermodynamic parameters of the Hayward regular BH in asymptotically anti-de Sitter, Minkowski, and de Sitter spacetimes, adopting both Euclidean path integral and Hamiltonian methods. Several other important works can be found in Refs. \cite{Cassani2024May,Sadeghi2024Jan,Aref'eva2024Feb,Abbas2024Apr, Ladghami2024May,Sokoliuk2024Jan,Yang2024Apr,Paul2024Apr,Ladghami2024May1,Ali2024Apr, Rakic2024Jun,Sadeghi2024Jan1,Davies2024Apr,Kruglov2024Feb,Ballesteros2024Feb, Yue2024Jul,Wu2020Jan,Wu2024Jun,Wu2023Jul,Wu2019Nov,Wu2020Aug}

It's implausible that BHs are completely isolated in the universe. It might be possible that they exist in dynamic and complex environments. In particular, there is a strong indication that supermassive BHs are the powerhouses behind active galactic nuclei \cite{Rees1984Sep,Kormendy1995Sep}. There is also substantial evidence indicating that dark matter envelopes most galaxies in a halo \cite{Bertone2018Oct}. Several important works based on BH-matter composite systems have been carried out for special types of dark-matter halo profiles. For instance, Xavier et al. studied shadows of non-rotating BH immersed in a Hernquist-like dark matter profile \cite{Xavier2023Mar}. Cardoso et al. discussed the effect of the Hernquist dark matter halo profile on the propagation of gravitational waves and on the geodesics \cite{Cardoso2022Mar}. Jusufi et al. \cite{Jusufi2019Aug} presented a new BH solution surrounded by a dark matter halo in the galactic center of the M87 galaxy from the universal rotation curve (URC) dark matter profile. Konoplya \cite{Konoplya2019Aug} considered a spherical model consisting of a Schwarzschild BH with a piecewise distribution of dark matter around it. In another work, Konoplya and Zhidenko \cite{Konoplya2022Jul} derived BH solutions in the presence of various dark matter halo profiles. Hou et al. \cite{Hou2018Jul} studied rotating BH at the centre of the Sgr* galaxy with cold dark matter and scalar field dark matter halos. Yang et al. \cite{Yang2024Jan} discussed the optical aspects of a rotating BH-dark matter system with a pseudo-isothermal dark matter halo profile. Liang et al. \cite{Liang2023Nov} discussed the thermodynamic aspects of a BH immersed in a perfect fluid dark matter halo. Carvalho \cite{Carvalho2023Dec} studied the thermodynamics of Einstein-Gauss-Bonnet (EGB) BH surrounded by three different distributions of dark matter halos. Anjum et al. \cite{Anjum2023May} studied Kerr BHs surrounded by perfect fluid dark matter and analysed the photon orbits, naked singularities and shadows with reference to observed results of the Event horizon telescope. Cappozziello et al. \cite{Capozziello2023May} studied the effects of dark matter spike near a supermassive black hole M87, in the framework of the Bumblebee Gravity. Jusufi \cite{Jusufi2023Feb} presented a spherically symmetric and asymptotically flat BH solution in which dark matter is made up of weakly interacting particles that orbit a supermassive black hole in the galactic centre, where the dark matter halo is formed by Einstein clusters. Pantig and \"{O}v\"{g}un \cite{Pantig2022May}, investigated the influence of different dark matter  halo profiles on the weak deflection angle. Some other interesting works related to BH systems in dark matter environments can be found in Refs. \cite{Stuchlik2021Nov,Pantig2023Jan,Ovgun2024Apr} 

The Dehnen density profile \cite{Dehnen1993Nov,Mo2010May} is commonly studied while dealing with dwarf galaxies, which often do not host BHs at their centers. However, recent observations suggest that massive BHs may also be present in these dwarf galaxies. Specifically, it was reported that a supermassive BH (SMBH) with a mass of around \(2.00 \times 10^5 M_{\odot}\) resides in the dwarf galaxy Mrk 462 \cite{BibEntry2024Jun}. Additionally, a study on BH-triggered star formation in the dwarf galaxy Henize 2-10 identified an SMBH with a mass of around \(1.00 \times 10^6 M_{\odot}\). Furthermore, a dynamical study of dark matter using photometric and spectroscopic data revealed the presence of a BH in Leo I, with a mass of \(3.3 \pm 2 \times 10^6 M_{\odot}\), accounting for 13$\%$ of the total mass of the Leo I galaxy \cite{Bustamante-Rosell2021Nov}.

Regarding the behaviour of null geodesics around BH spacetimes, there are several important works carried out in different aspects. For instance, recently, Virbhadra derived a novel formula for the compactness of a gravitational lens, allowing for accurate upper bounds through observation of relativistic images without any prior knowledge of the lens mass or distances \cite{Virbhadra2022Apr}. Modification or corrections to usual Schwarschild spacetime offers interesting possibilities; for instance, Battista \cite{Battista2024Jan} explored the Schwarzschild geometry within the formulation of an effective field theory, deriving the quantum corrected metric. He also analysed the metric horizons, geodesics, and energy-extraction mechanism violating the null energy condition. Battista and Esposito \cite{Battista2022Dec} also investigated the geodesic motion in Euclidean Schwarzschild geometry within the equatorial plane and provided explicit forms using incomplete elliptic integrals. They presented the possibility of the absence of elliptic-like orbits and that only unbounded first-kind orbits are allowed, which are in contrast with GR, where unbounded second-kind orbits are also allowed. Other recent related works can be found in Refs. \cite{Nekouee2024Apr,Majumder2024Jan,Wang2024Mar,Javed2024Jul, Boshkayev2024Jan,Pantig2024Jul}.
 
In this paper, we study the influence of the Dehnen dark matter profile on a Schwarszchild BH. To achieve this, we begin by deriving a new BH metric that incorporates the Dehnen profile and then study the aforementioned aspects. More specifically, in the first part of the paper, we study the thermodynamical aspects of the BH-dark matter system and how the dark matter core density affects the thermodynamical stability of such a system. In the second part of the paper, we study the null geodesics of the BH-dark matter system and the stability of the photon orbit with respect to the dark matter parameters. The paper is organised as follows: In Section \ref{sec2} we discuss the theoretical framework of the Dehnen-type dark matter profile where we obtain the effective metric function of the BH-dark matter profile. In Section \ref{sec3}, we discuss the thermodynamic aspects and the stability of our system. In Section \ref{sec4}, we obtain the effective potential of our BH system and study the null geodesics and their stability through dynamical systems and Lyapunov exponents and studied if the circular geodesics exhibit chaotic behaviour using the chaos-bound condition. Finally, in Section \ref{conc} we conclude with the results of our study.

\section{Dark Matter Density Profile and Metric Function}
\label{sec2}
In this paper, we intend to investigate the derivation of a non-rotating and uncharged BH solution in the vicinity of a Dehnen-type dark matter halo. The density profile of the Dehnen dark matter halo is a special case of a double power-law profile given by \cite{Mo2010May}
\begin{equation}
\rho_{dm} = \rho_s \left(\frac{r}{r_s}\right)^{-\gamma } \left[\left(\frac{r}{r_s}\right)^{\alpha }+1\right]^{\frac{\gamma -\beta }{\alpha }}
\label{doub_pow_law}
\end{equation}
We consider one of several possible Dehnen profiles, where $\gamma$ determines the specific variant of the profile. Some of the allowed variants of the Dehnen profile can be obtained from setting $(\alpha,\beta,\gamma) = (1, 4, \gamma)$. The values of $\gamma$ lies within $[0,3]$. For instance, $\gamma = 3/2$ has been used to fit the surface brightness profiles of elliptical galaxies which closely resembles the de Vaucouleurs $r^{1/4}$ profile \cite{Shakeshaft2012Dec}. Recently, Pantig and \"{O}v\"{g}un \cite{Pantig2022Aug} studied the effect of Dehnen dark matter halo in an ultrafaint dwarf galaxy. In this work, we use the parameters $(\alpha, \beta, \gamma) = (1, 4, 0)$. This gives
\begin{equation}
\rho_{D} = \frac{\rho_s}{\left(\frac{r}{r_s}+1\right)^4}
\label{dehnen_profile}
\end{equation}
where $\rho_s$ and $r_s$ denote the central halo density and the halo core radius respectively.

Let us now obtain the mass distribution of the dark matter profile. The mass profile can be calculated through the relation
\begin{equation}
M_{D} = \int_0^r 4\pi \rho_D (r')r'^2 dr' = \frac{4 \pi  r^3 r_s^3 \rho _s}{3 \left(r_s+r\right)^3}
\label{mass_prof}
\end{equation}
The tangential velocity of a test particle moving in the dark matter halo can be determined from the mass distribution of the halo profile in a spherically symmetric spacetime. In units of $G = c = 1$, the tangential velocity is directly related to the mass profile as
\begin{equation}
v_{D}^2 = \frac{M_D}{r} = \frac{4 \pi  r^2 r_s^3 \rho _s}{3 \left(r_s+r\right)^3},
\label{velc1} 
\end{equation}
A spherically symmetric line element describing a pure dark matter halo can be assumed as 
\begin{equation}
ds^2 = -\mathcal{F}(r) dt^2 + \mathcal{G}(r)^{-1} dr^2 + r^2 (d\theta^2 + \sin^2 \theta d\phi^2),
\label{line_element_dm}
\end{equation}
where $\mathcal{F}(r)$ and $\mathcal{G}(r)$ represents the redshift function and the shape function respectively. There exists a close relationship between the redshift function $F(r)$ and the tangential velocity \cite{Yang2024Jan}
\begin{equation}
v_D^2 = r \frac{d}{dr} \ln \sqrt{\mathcal{F}(r)}.
\label{vel_redshift}
\end{equation}
In this work, we shall work with the setting that the redshift function and the shape function are equal, i.e. $\mathcal{F}(r) = \mathcal{G}(r)$.
Using Eqs. \eqref{velc1} and \eqref{vel_redshift} we obtain 
\begin{equation}
\mathcal{F}(r) = \mathcal{G}(r) = \exp \left(-\frac{4 \pi  r_s^3 \left(r_s+2 r\right) \rho _s}{3 \left(r_s+r\right)^2}\right) \approx 1 -\frac{4 \pi  r_s^3 \left(r_s+2 r\right) \rho _s}{3 \left(r_s+r\right)^2}
\label{Fr}
\end{equation}
where we retained the leading order terms of the equation.

The Einstein field equations need to be satisfied by the spacetime 
\eqref{Fr} is 
\begin{equation}
R_{\mathcal{A} \mathcal{B}} - \frac{1}{2} R g_{\mathcal{A} \mathcal{B}} = \kappa^2 T_{\mathcal{A} \mathcal{B}} (D),
\label{EFE}
\end{equation}
where $g_{\mathcal{A} \mathcal{B}}$, $R_{\mathcal{A} \mathcal{B}}$ and $R$ denotes the metric tensor, Ricci tensor and the Ricci scalar respectively. Also $T_{\mathcal{A} \mathcal{B}} (D)$ denotes the energy-momentum tensor of the Dehnen dark matter halo spacetime, which can be expressed as $T_{\mathcal{A}}^{\mathcal{B}} = g^{\mathcal{B} \mathcal{C}} T_{\mathcal{A} \mathcal{C}} = \text{diag} [-\rho, p_r, p, p]$. Thus solving the field equations one obtains, 
\begin{equation}
\begin{aligned}
\kappa^2 T_t^{t (D)}= & \mathcal{G}(r)\left(\frac{1}{r} \frac{\mathcal{G}^{\prime}(r)}{\mathcal{G}(r)}+\frac{1}{r^2}\right)-\frac{1}{r^2}, \\
\kappa^2 T_r^{r (D)} &= \mathcal{G}(r)\left(\frac{1}{r^2} + \frac{1}{r} \frac{\mathcal{F}^{\prime}(r)}{\mathcal{F}(r)}\right) - \frac{1}{r^2}, \\
\kappa^2 T_\theta^{\theta (D)} &= \kappa^2 T_\phi^{\phi (D)} = \frac{1}{2} \mathcal{G}(r) \left( \frac{\mathcal{F}^{\prime \prime}(r) \mathcal{F}(r) - \mathcal{F}^{\prime 2}(r)}{\mathcal{F}^2(r)} + \frac{1}{2} \frac{\mathcal{F}^{\prime 2}(r)}{\mathcal{F}^2(r)} \right. \\& \left.\hspace{2cm}+ \frac{1}{r} \left( \frac{\mathcal{F}^{\prime}(r)}{\mathcal{F}(r)} + \frac{\mathcal{G}^{\prime}(r)}{\mathcal{G}(r)} \right) + \frac{\mathcal{F}^{\prime}(r) \mathcal{G}^{\prime}(r)}{2 \mathcal{F}(r) \mathcal{G}(r)} \right) .
\end{aligned}
\end{equation}
For our convenience, for the combined system of the Schwarzschild BH and the dark matter halo, let us now rewrite the metric functions as follows
\begin{equation}
\begin{aligned}
& f(r)= \mathcal{F}(r) + \mathcal{F}_1(r), \\
& g(r)= \mathcal{G}(r)+\mathcal{F}_2 (r).
\end{aligned}
\end{equation}
where $\mathcal{F}_1 (r)$ and $\mathcal{G}_2 (r)$ are unknown functions that can be determined from the BH parameters and the dark matter halo parameters. Thus the combined spacetime metric that contains the contribution of both the BH and dark matter spacetime can be expressed as 
\begin{equation}
ds^2 = - f(r) dt^2 + g(r)^{-1} dr^2 + r^2 (d\theta^2 + \sin^2 \theta d\phi^2),
\label{com_met}
\end{equation}
Given this, the Einstein field equation can now be written as
\begin{equation}
R_{\mathcal{A} \mathcal{B}} - \frac{1}{2} R g_{\mathcal{A} \mathcal{B}} = \kappa^2 \left[T_{\mathcal{A} \mathcal{B}} (D) + T_{\mathcal{A} \mathcal{B}} (BH)\right],
\label{EFEcom}
\end{equation} 
where $T_{\mathcal{A} \mathcal{B}} (BH)$ arises from the matter content of the pure BH spacetime.

\begin{figure*}
\centerline{\includegraphics[scale=0.5]{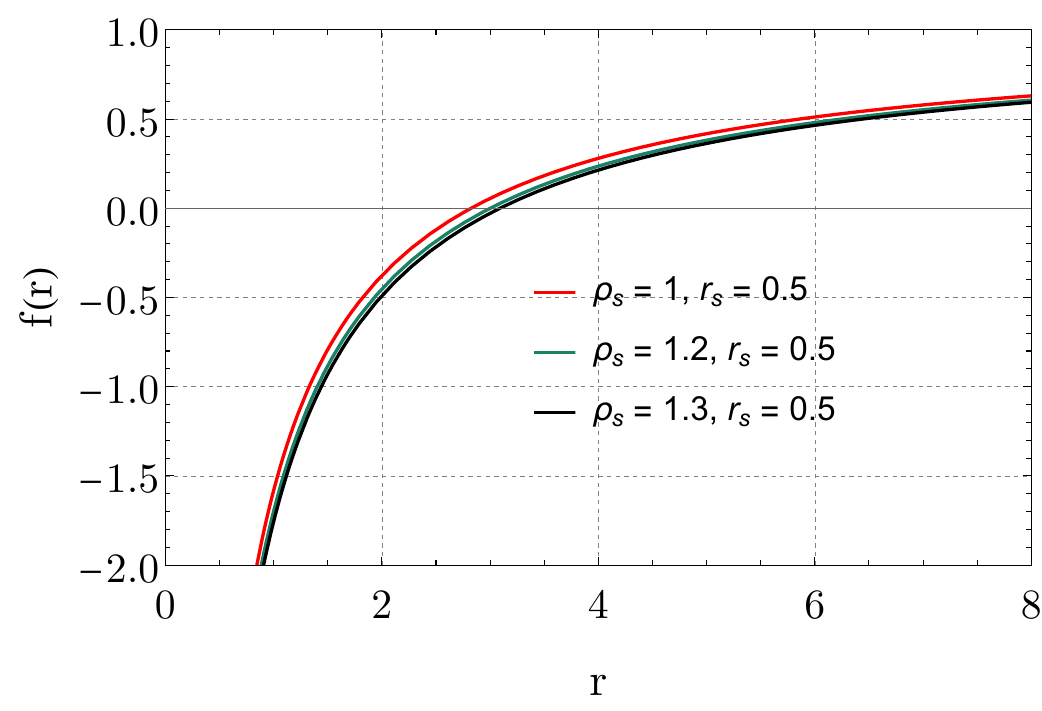}}
\caption{The metric function is plotted for different values of the dark matter central density $\rho_s$ by setting the core radius $r_s$ fixed. The black dot represents the extremal point.}
\label{frplot}
\end{figure*}

Using the combined space-time metric \eqref{com_met} and Einstein field equations \eqref{EFEcom} one yields
\begin{equation}
\begin{aligned}
& \left(\mathcal{G}(r)+ \mathcal{F}_2(r)\right)\left(\frac{1}{r^2}+\frac{1}{r} \frac{\mathcal{G}^{\prime}(r)+\mathcal{F}_2^{\prime}(r)}{\mathcal{G}(r)+ \mathcal{F}_2(r)}\right) = \mathcal{G}(r)\left(\frac{1}{r^2}+\frac{1}{r} \frac{\mathcal{G}^{\prime}(r)}{\mathcal{G}(r)}\right), \\
& \left(\mathcal{G}(r)+ \mathcal{F}_2(r)\right)\left(\frac{1}{r^2}+\frac{1}{r} \frac{\mathcal{F}^{\prime}(r)+\mathcal{F}_1^{\prime}(r)}{\mathcal{F}(r)+\mathcal{F}_1(r)}\right)= \mathcal{G}(r)\left(\frac{1}{r^2}+\frac{1}{r} \frac{\mathcal{F}^{\prime}(r)}{\mathcal{F}(r)}\right) .
\end{aligned}.
\label{field_eqs}
\end{equation}
Now using the Schwarzschild BH as the boundary condition, one obtains the solutions to the above differential Eqs. \eqref{field_eqs} as
\begin{equation}
\begin{aligned}
\mathcal{F}_2 (r) &= - \frac{2M}{r} \\
\mathcal{F}_1 (r) &= \exp \left[\int \frac{\mathcal{G}(r)}{\mathcal{G}(r) + \mathcal{F}_2 (r)}\left(\frac{1}{r}+\frac{\mathcal{F}^{\prime}(r)}{\mathcal{F}(r)}\right) d r - \frac{1}{r}dr\right] - \mathcal{F}(r).
\end{aligned}
\end{equation}
As stated earlier, we have assumed $\mathcal{F}(r) = \mathcal{G}(r)$. This results into $\mathcal{F}_1(r) = \mathcal{F}_2(r) = -\frac{2M}{r}$. This in turn gives $f(r) = g(r) = \mathcal{F}(r) - \frac{2M}{r}$. Therefore the final form of the metric in the presence of both the BH and the Dehnen dark matter halo becomes 
\begin{equation}
ds^2 = -f(r) dt^2 + f(r)^{-1} dr^2 + r^2 ( d\theta^2 + \sin^2 \theta d\phi^2),
\label{fin_metric} 
\end{equation}
with \begin{equation} 
f(r) = 1 - \frac{2M}{r} - \frac{4 \pi  r_s^3 \left(r_s+2 r\right) \rho _s}{3 \left(r_s+r\right)^2}.
\label{fin_metric_funct}
\end{equation}
This is the effective metric function of our composite dark matter-BH system. Fig. \ref{frplot} represents the plot of the metric function \eqref{fin_metric_funct} as a function of $ r $. It is evident from the plot that the central density of the dark matter halo significantly influences the existence of BH horizons while keeping the core radius fixed at \( r_s = 0.5 \). The horizons can be determined by the condition \( f(r) = 0 \). We see that there exists a unique horizon for each of the combinations of the dark matter halo parameters $\rho_s$ and $r_s$. 
\subsection{Singularity Test}
It is known that the Schwarszchild BH possess a true physical singularity at $r = 0$ and coordinate singularity at $r = 2m$. Let us now investigate whether the presence of the dark matter halo influences the existence of singularity or does it plays any role in making the BH regular at $r = 0$. To do so, let us calculate the curvature invariants, i.e Ricci scalar $(R)$, Ricci squared ($R_{\mu \nu} R^{\mu \nu}$) and Kretschmann scalar ($K$). For the metric \eqref{fin_metric}, they can be written as
\begin{equation}
R=-\frac{r^2 f^{\prime \prime}(r)+4 r f^{\prime}(r)+2 f(r)-2}{r^2},
\label{ricc1}
\end{equation}
\begin{equation}
R_{\mu \nu} R^{\mu \nu}=\frac{r^4 f^{\prime \prime}(r)^2+8 r^2 f^{\prime}(r)^2+8 f(r)\left(r f^{\prime}(r)-1\right)+4 r f^{\prime}(r)\left(r^2 f^{\prime \prime}(r)-2\right)+4 f(r)^2+4}{2 r^4}, 
\label{rsq}
\end{equation}
and 
\begin{equation}
K = R_{\mu \nu \alpha \beta} R^{\mu \nu \alpha \beta}=f^{\prime \prime}(r)^2+\frac{4 f^{\prime}(r)^2}{r^2}+\frac{4(f(r)-1)^2}{r^4}.
\label{kretc}
\end{equation}
 Using Eq. \eqref{fin_metric_funct}, Eqs. \eqref{ricc1}, \eqref{rsq} and \eqref{kretc} can be found as 
\begin{equation}
\begin{aligned}
&R = \frac{8 \pi  r_s^5 \left(r_s+4 r\right) \rho _s}{3 r^2 \left(r_s+r\right)^4}, \\
&R_{\mu \nu} R^{\mu \nu} = \frac{32 \pi ^2 r_s^8 \left(18 r^4+24 r^3 r_s+22 r^2 r_s^2+8 r r_s^3+r_s^4\right) \rho _s^2}{9 r^4 \left(r_s+r\right){}^8} \\
&K =R_{\mu \nu \sigma \rho}\,R^{\mu \nu \sigma \rho} = \frac{48 M^2}{r^6} + \\ & \frac{64 \pi  r_s^3 \rho _s \left(3 M \left(6 r^3+6 r^2 r_s+4 r r_s^2+r_s^3\right) \left(r_s+r\right){}^4+\pi  r r_s^3 \left(12 r^6+24 r^5 r_s+46 r^4 r_s^2+44 r^3 r_s^3+26 r^2 r_s^4+8 r r_s^5+r_s^6\right) \rho _s\right)}{9 r^5 \left(r_s+r\right)^8} 
\end{aligned}
\label{curv_scl}
\end{equation}
\begin{figure*}
\centerline{\includegraphics[scale=0.55]{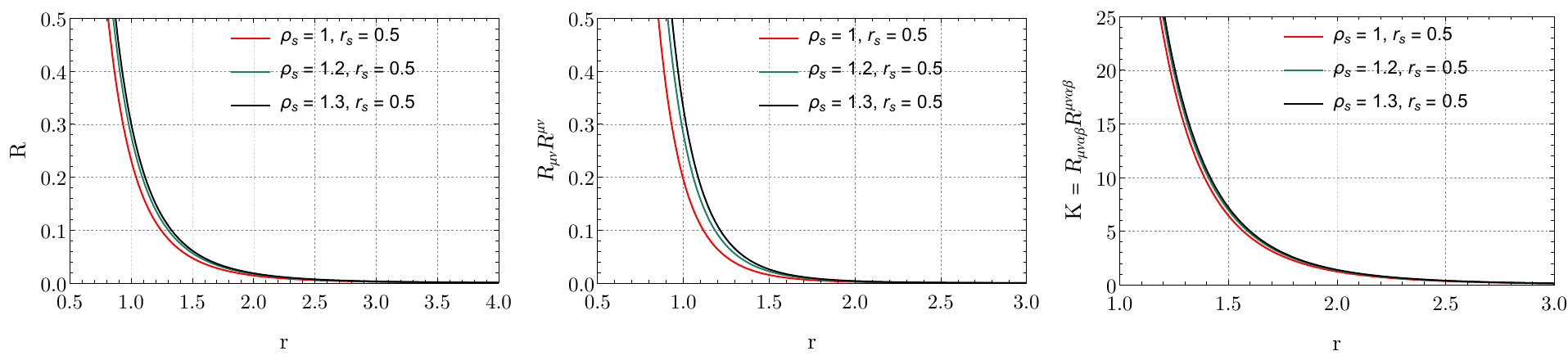}}
\caption{The Ricci scalar, Ricci squared and the Kretschmann scalar are shown for different combinations of dark matter parameters.}
\label{scalar_fig}
\end{figure*}
It is seen from Eq. \eqref{curv_scl} as well as from Fig. \ref{scalar_fig}  that the curvature scalars $R$, $R_{\mu\nu}R^{\mu \nu}$ and $K$ all diverge at $r = 0$, which signifies that the BH is not regular at $r = 0$. Therefore, the presence of the dark matter halo background does not affect the occurrence of a physical singularity at $r = 0$. It can also be seen from these equations that as $r \to \infty$, the scalars vanish asymptotically. This represents that the BH solution in our model is unique and there exists a physical singularity at $r = 0$. A similar observation was made by Zhang et al in the case of a Bardeen BH surrounded by perfect fluid dark matter \cite{Zhang2021May}. 

\section{Thermodynamical Parameters}
\label{sec3}
In this section, we shall obtain the thermodynamic functions relevant to the BH system in the presence of the Dehnen dark matter halo to see how the halo parameters affect them. At first, we obtain the Hawking temperature $T_H$ that can be obtained through the direct relation with the metric function Eq.  \eqref{fin_metric_funct} which is 
\begin{equation}
T = \left.\frac{f'(r)}{4\pi} \right|_{r=r_h} = \frac{-12 \pi  r_h r_s^4 \rho _s+9 r_h r_s^2+9 r_h^2 r_s+3 r_h^3-4 \pi  r_s^5 \rho _s+3 r_s^3}{12 \pi  r_h \left(r_h+r_s\right)^3},
\label{temp}
\end{equation}
where $r_h$ is the horizon radius.
\begin{figure*}
\centerline{\includegraphics[scale=0.5]{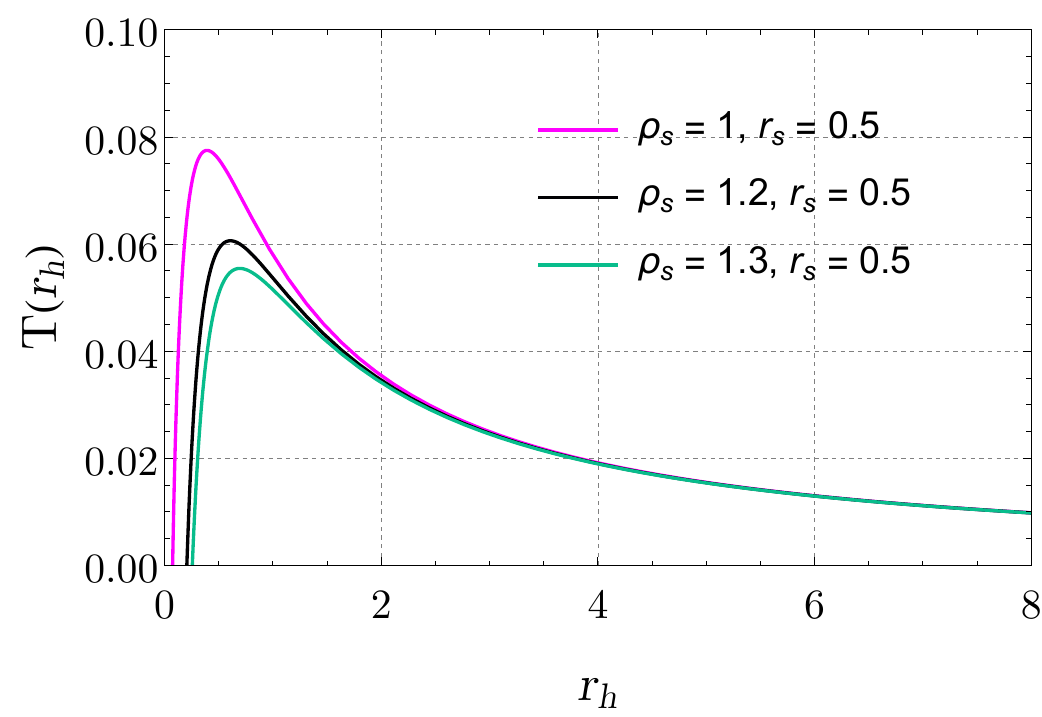}}
\caption{The Hawking temperature is plotted for different values of the dark matter central density $\rho_s$ by setting the core radius $r_s$ fixed.}
\label{temp_plot}
\end{figure*}
Fig. \ref{temp_plot} illustrates the behaviour of the Hawking temperature $T$ as a function of the horizon radius \( r_h \) for different values of the dark matter central core density \( \rho_s \). The plots show that the temperature initially rises with increasing horizon radius, reaches a maximum value, and subsequently decreases as the horizon radius continues to increase. Additionally, it is observed that the peak value of the Hawking temperature diminishes with increasing central core density. Therefore, it is reasonable to state that for a fixed value of the horizon radius within a physically viable range, BHs surrounded by a dark matter halo with a lower central core density exhibit higher temperatures. It is equally significant that the occurrence of the peak for higher values of $\rho_s$ attenuates. Therefore, for larger BHs, it can be speculated that there shall be no abrupt phase transition corresponding to higher values of $\rho_s$. Moreover, the horizon radius at which the peak temperature occurs shifts to larger values as the central core density of the dark matter halo increases, and as the horizon radius increases further, the temperature asymptotically vanishes eventually.

Again, setting $f(r_h) = 0$ leads us to obtain the Arnowitt-Deser-Misner (ADM) mass ($M$) of the BH given by
\begin{equation}
M = \frac{1}{2} r_h \left(1-\frac{4 \pi  r_s^3 \rho _s \left(2 r_h+r_s\right)}{3 \left(r_h+r_s\right)^2}\right),
\label{mass}
\end{equation}
Clearly, in the limit of vanishing $\rho_s$, the ADM mass reduces to the mass of Schwarszchild BH 
\begin{equation}
M_{sch} = \frac{r_h}{2},
\label{sch_mass} 
\end{equation}
\begin{figure*}
\centerline{\includegraphics[scale=0.5]{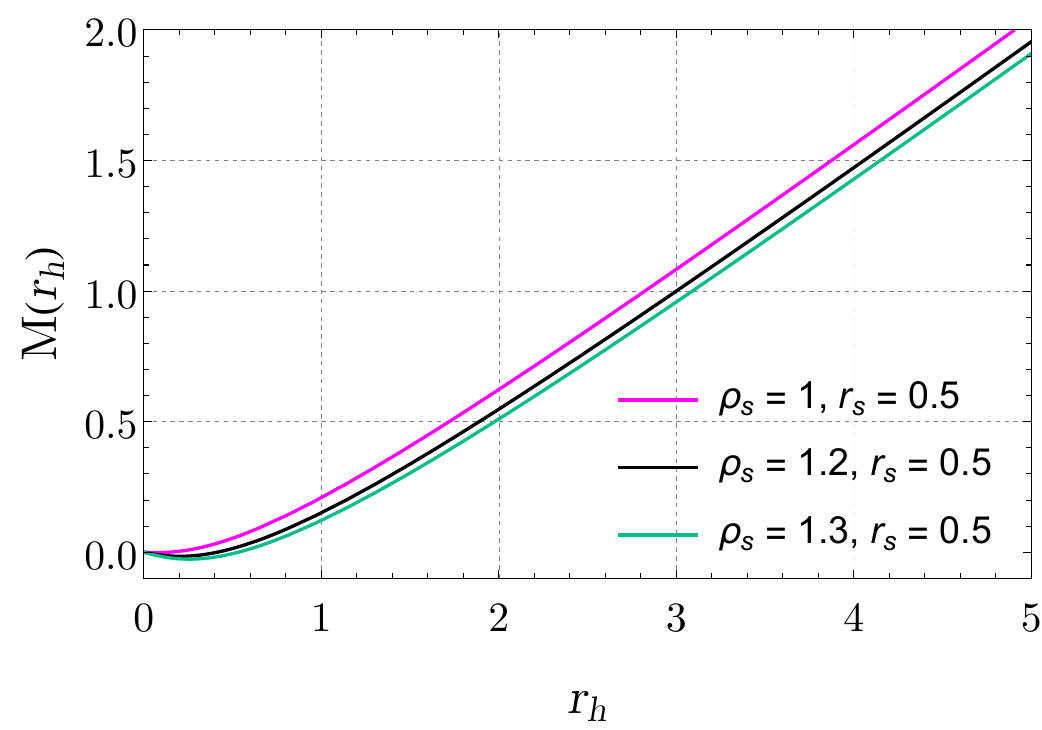}}
\caption{The mass is plotted for different values of the dark matter central density $\rho_s$ by setting the core radius $r_s$ fixed.}
\label{M_rh_plot}
\end{figure*}
From Fig. \ref{M_rh_plot}, we observe that the ADM mass increases monotonically with increasing horizon radius. 
The entropy of the BH can be calculated as
\begin{equation}
S_{BH} = \int \frac{dM}{T} = \pi r_h^2,
\label{bh_entropy}
\end{equation}
One notices that the dark matter halo distribution does not affect the entropy of the event horizon. 

To study the local and global thermodynamic stability of BH systems, it is essential to examine thermodynamic parameters like specific heat capacity and Helmholtz free energy respectively.  
The specific heat is obtained as
\begin{equation}
C = \frac{\partial M}{\partial T} = -\frac{2 \pi  r_h^2 \left(r_h+r_s\right) \left(12 \pi  r_h r_s^4 \rho _s-9 r_h r_s^2-9 r_h^2 r_s-3 r_h^3+4 \pi  r_s^5 \rho _s-3 r_s^3\right)}{18 r_h^2 r_s^2 \left(2 \pi  r_s^2 \rho _s-1\right)+4 r_h r_s^3 \left(4 \pi  r_s^2 \rho _s-3\right)-12 r_h^3 r_s-3 r_h^4+r_s^4 \left(4 \pi  r_s^2 \rho _s-3\right)},
\label{spheat}
\end{equation}
\begin{figure*}
\centerline{\includegraphics[scale=0.6]{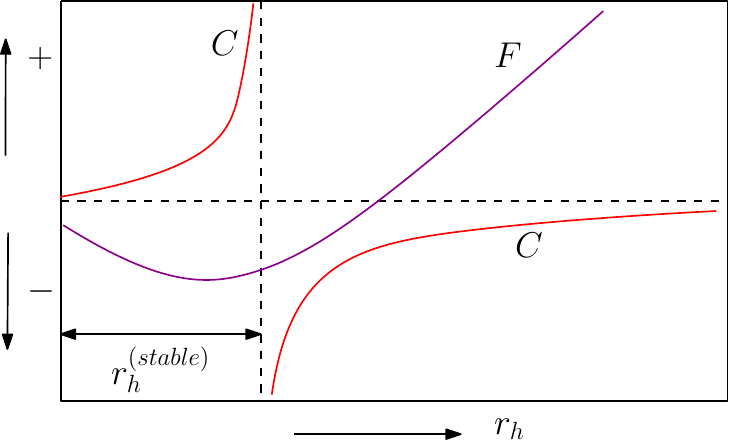}}
\caption{A toy model: Illustration of simultaneous local and global thermodynamic stability. Here $r_h^{(stable)}$ represents the range where the BH is both locally and globally stable.}
\label{toy}
\end{figure*}
\begin{figure*}
\hspace{-0.5cm}
\centerline{\includegraphics[scale=0.45]{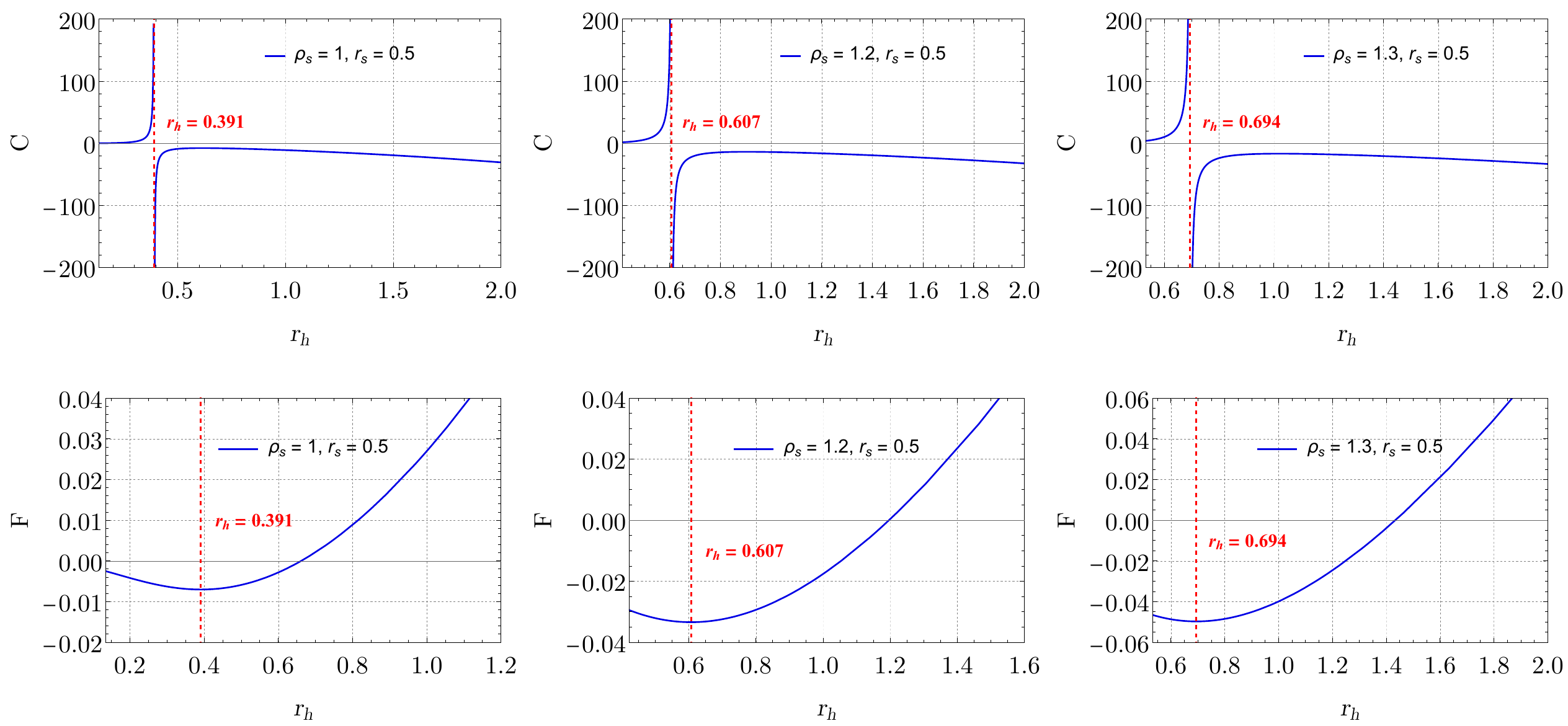}}
\caption{The specific heat and free energy are plotted for different values of the dark matter central density $\rho_s$ by setting the core radius $r_s$ fixed at $0.5$. The red dashed line in each case shows the upper limit of the horizon radius within which simultaneous local and global thermodynamic stability is seen.}
\label{spec_free}
\end{figure*}
The sign of the specific heat capacity offers conducive information about the system's local thermodynamic stability. A positive specific heat capacity suggests that the BH system is thermodynamically stable, which indicates that it requires energy to increase its temperature. In contrast, a negative specific heat capacity indicates thermodynamic instability, in which the BH system absorbs heat while paradoxically cooling down, suggesting a peculiar response to energy absorption. By numerically plotting the specific heat $C$ with the horizon radius $r_h$ as shown in Fig. \ref{spec_free}, we can infer how the dark matter core density parameter $\rho_s$ affects the second-order phase transition. The specfic heat diverges at the point of peak temperature, also known as Davies' point. This point separates the thermodynamically stable, positive specific heat region from the thermodynamically unstable, negative specific heat region. As $\rho_s$ increases, the Davies' point shifts towards the right, indicating an increase in the thermodynamically stable region of the BH. Concerning $\rho_s$, we notice that for lower values, the second-order phase transition occurs at smaller values of the critical horizon radius $r_h^c$. This phase transition causes the BH to shift from a locally stable state to an unstable one. Therefore, at a horizon radius $r_h < r_h^c$, the BH is locally stable, whereas it becomes unstable for $r_h > r_h^c$. From the plots, it is visually clear that as $\rho_s$ increases, the range of the horizon radius for which the BH is locally stable diminishes. In other words, for larger BHs, the specific heat is always negative, which suggests that larger BHs are locally unstable. 

Conversely, to examine the global stability of the BH system, we analyze the Helmholtz free energy ($F$) in our model. A positive value of $F$ indicates the global instability of the thermodynamic system, whereas a negative value indicates global stability. The Helmholtz free energy at the horizon can be calculated as
\begin{equation}
F = M - TS = \frac{r_h \left(r_h^2 \left(9 r_s-16 \pi  r_s^3 \rho _s\right)+r_h \left(9 r_s^2-12 \pi  r_s^4 \rho _s\right)+3 r_h^3-4 \pi  r_s^5 \rho _s+3 r_s^3\right)}{12 \left(r_h+r_s\right)^3}
\label{free_energy}
\end{equation}
From Fig. \ref{spec_free}, it is clear that the free energy for our BH system exhibits a flip in sign at some particular value of horizon radius depending on the value of $\rho_s$. The point of phase transition (Hawking-Page point) occurs at a larger horizon radius for larger values of $\rho_s$.  The red dashed lines in the $F$ vs $r_h$ plot also represent the upper limit of physically stable BH states. The existence intervals of physically allowed horizon radius corresponding to local and global stability are listed in Tab. \ref{tab_stab}. 
\begin{table*}[htb]
\centering
\begin{tabular}{||c|c|c|c||}
\hline
$\rho_s$ & Local Stability          & Global Stability         & Local + Global Stability \\ \hline
$1$      & $r_h \in [0.135, 0.391]$ & $r_h \in [0.135, 0.661]$ & $r_h \in [0.135, 0.391]$ \\ \hline
$1.2$    & $r_h \in [0.412, 0.607]$ & $r_h \in [0.412, 1.196]$ & $r_h \in [0.412, 0.607]$ \\ \hline
$1.3$    & $r_h \in [0.531, 0.694]$ & $r_h \in [0.531, 1.433]$ & $r_h \in [0.531, 0.694]$ \\ \hline
\end{tabular}
\caption{Existence intervals of BH states corresponding to local and global thermodynamic stability.}
\label{tab_stab}
\end{table*}
\section{Effective Potential and Null Geodesics}
\label{sec4}
Null geodesics are the paths taken by massless particles like photons in spacetime. Mathematically, these trajectories are represented by the condition $ds^2 = 0$. The trajectories of massless particles may play an important role in estimating the effect of dark matter surrounding a BH system. The presence of dark matter can alter the spacetime curvature, leading to the deviation of null geodesics. To investigate it further, we need to solve the set of geodesic equations, which is carried out as follows:

To calculate the geodesic equations of motion, one may start from the general form of a spherically symmetric metric given as
\begin{equation}
ds^2 = - A(r)^2 dt^2 + B(r)^2 dr^2 + r^2 d\theta^2 + r^2 \sin^2 \theta d\phi^2,
\label{gen_met} 
\end{equation}
The metric \eqref{gen_met} possesses time translation and spherical symmetry, which implies that the Killing vector associated with them gives a conserved quantity along the geodesics, given by 
\begin{equation}
K_\mu \dot{x}^\mu = \text{ constant}.
\label{kill}
\end{equation}
Here the $`.'$ represents the derivative with respect to the affine parameter $\lambda$. The Killing vectors related to the time translational symmetry and spherical symmetry are given as
\begin{equation}
K_\mu = ( - A(r), 0, 0, 0),
\label{time_killing}
\end{equation}
and 
\begin{equation}
K_\mu = (0, 0, 0, r^2 \sin^2 \theta),
\label{sph_killing}
\end{equation} respectively.
From Eqs. \eqref{gen_met} and \eqref{time_killing}, we obtain the E-Equation
\begin{equation}
E = A(r)\dot{t} = \text{constant}.
\label{E-eqn}
\end{equation}
Keeping in mind that we are dealing with spherical symmetry, we have the liberty to restrict the observer to the equatorial plane by setting $\theta = \pi/2$. This allows us to obtain the L-Equation from Eqs. \eqref{gen_met} and \eqref{sph_killing} as
\begin{equation}
L = r^2 \dot{\phi} = \text{constant},
\label{L-eqn}
\end{equation}
One can also find that the norm of the tangent vector to the geodesic is a conserved quantity, which implies
\begin{equation}
\epsilon = - g_{\mu \nu} \dot{x}^\nu \dot{x}^\mu,
\label{eps}
\end{equation}
where $\epsilon = 0$ for null geodesics. Using the metric \eqref{gen_met} in Eq. \eqref{eps} we get
\begin{equation}
-\epsilon = -A(r) \dot{t}^2 + B(r) \dot{r}^2 + r^2 \dot{\phi}^2.
\label{eps1}
\end{equation}
Eq. \eqref{eps1} can also be written as
\begin{equation}
\dot{r}^2 = \frac{E^2}{A(r) B(r)} - \frac{L^2}{r^2 B(r)} - \frac{\epsilon}{B(r)}.
\label{rdotsq}
\end{equation}
To find the geodesic equations we utilise the Lagrangian given by 
\begin{equation}
\mathcal{L} = \frac{1}{2}g_{\mu \nu} \dot{x}^\alpha \dot{x}^\beta = \frac{1}{2}\left(-A(r) \dot{t}^2 + B(r) \dot{r}^2 + r^2 \dot{\phi}^2\right).
\label{Lag}
\end{equation}
Using the Euler-Lagrange Equation in the $r$-coordinate
\begin{equation}
\frac{d}{d\lambda}\left( \frac{\partial \mathcal{L}}{\partial \dot{r}}\right) = \frac{\partial \mathcal{L}}{\partial r}.
\label{Lagr}
\end{equation}
This gives
\begin{equation}
\dot{p}_r = \frac{1}{2}\left(- \frac{\partial A(r)}{\partial r}\dot{t}^2 + \frac{\partial B(r)}{\partial r}\dot{r}^2 + 2r \dot{\phi}^2 \right).
\label{prdot}
\end{equation}
using the conjugate momentum in $r$-coordinate 
\begin{equation}
p_r = \frac{\partial \mathcal{L}}{\partial \dot{r}}= \dot{r}B(r)
\label{pr}
\end{equation}
Thus by using Eqs. \eqref{E-eqn}, \eqref{L-eqn}, \eqref{prdot} and \eqref{pr} we can obtain the set of equations of motion for null-geodesics in the general spherically symmetric spacetime \eqref{gen_met} as:
\begin{equation}
\begin{aligned}
\dot{t} &= E A(r)^{-1} \\
\dot{\phi} &= \frac{L}{r^2} \\
\dot{r} &= p_r B(r)^{-1} \\
\dot{p}_r &= \frac{1}{2}\left(- \frac{E^2}{A(r)^2} \frac{\partial A(r)}{\partial r} + \frac{p_r^2}{B(r)^2} \frac{\partial B(r)}{\partial r} + \frac{2L^2}{r^3}\right).
\end{aligned}
\label{geod_eqs}
\end{equation}

From Eq. \eqref{rdotsq}, we find
\begin{equation}
\frac{1}{2}\dot{r}^2 + V_{eff} = \frac{E^2}{2},
\label{eff1}
\end{equation}
where \begin{equation}
V_{eff} = - \frac{E^2}{2A(r) B(r)} + \frac{L^2}{2r^2} B(r)^{-1} + \frac{\epsilon}{B(r)}.
\label{eff2}
\end{equation}
In the present work, we have shown in the last section that the effective metric including both the Schwarzszchild BH and the dark matter is given by Eq. \eqref{fin_metric} with the metric function \eqref{fin_metric_funct}. Comparing the general metric \eqref{gen_met} with the metric \eqref{fin_metric} we see $A(r) = f(r)$, $B(r) = f(r)^{-1}$. Thus the system of Eqs. \eqref{geod_eqs} becomes
\begin{equation}
\begin{aligned}
\dot{t} &= E \left(1 - \frac{2M}{r} -\frac{4 \pi  r_s^3 \rho _s \left(2 r_h+r_s\right)}{3 \left(r_h+r_s\right){}^2}\right)^{-1} \\
\dot{\phi} &= \frac{L}{r^2} \\
\dot{r} &= p_r \left(1 - \frac{2M}{r} -\frac{4 \pi  r_s^3 \rho _s \left(2 r_h+r_s\right)}{3 \left(r_h+r_s\right)^2}\right) \\
\dot{p}_r &= \frac{L^2}{r^3}-\frac{p_r^2 \left(3 M \left(r_s+r\right)^3+4 \pi  r^3 r_s^3 \rho _s\right)}{3 r^2 \left(r_s+r\right)^3} \\& \hspace{3cm}-\frac{3 E^2 \left(r_s+r\right) \left(3 M \left(r_s+r\right)^3+4 \pi  r^3 r_s^3 \rho _s\right)}{\left(6 M \left(r_s+r\right)^2+r \left(-3 r^2+4 \pi  r_s^4 \rho _s+8 \pi  r r_s^3 \rho _s-3 r_s^2-6 r r_s\right)\right)^2}.
\end{aligned}
\label{geod_eqs_mod}
\end{equation}
The effective potential for our model becomes
\begin{equation}
V_{eff} = \frac{L^2 \left(-\frac{2 M}{r}-\frac{4 \pi  \left(r_s+2 r\right) r_s^3 \rho _s}{3 \left(r_s+r\right){}^2}+1\right)}{2 r^2}
\label{eff_pot}
\end{equation}
where we have set $\epsilon = 0$ as we are dealing with null geodesics. Moreover, $E$ is set to be zero for plotting purposes, as the presence of $E$ is merely responsible for the shift in the effective potential amplitude and does not affect the photon sphere radii of the circular orbits. 
\begin{figure*}[tbh]
\centerline{\includegraphics[scale=0.3]{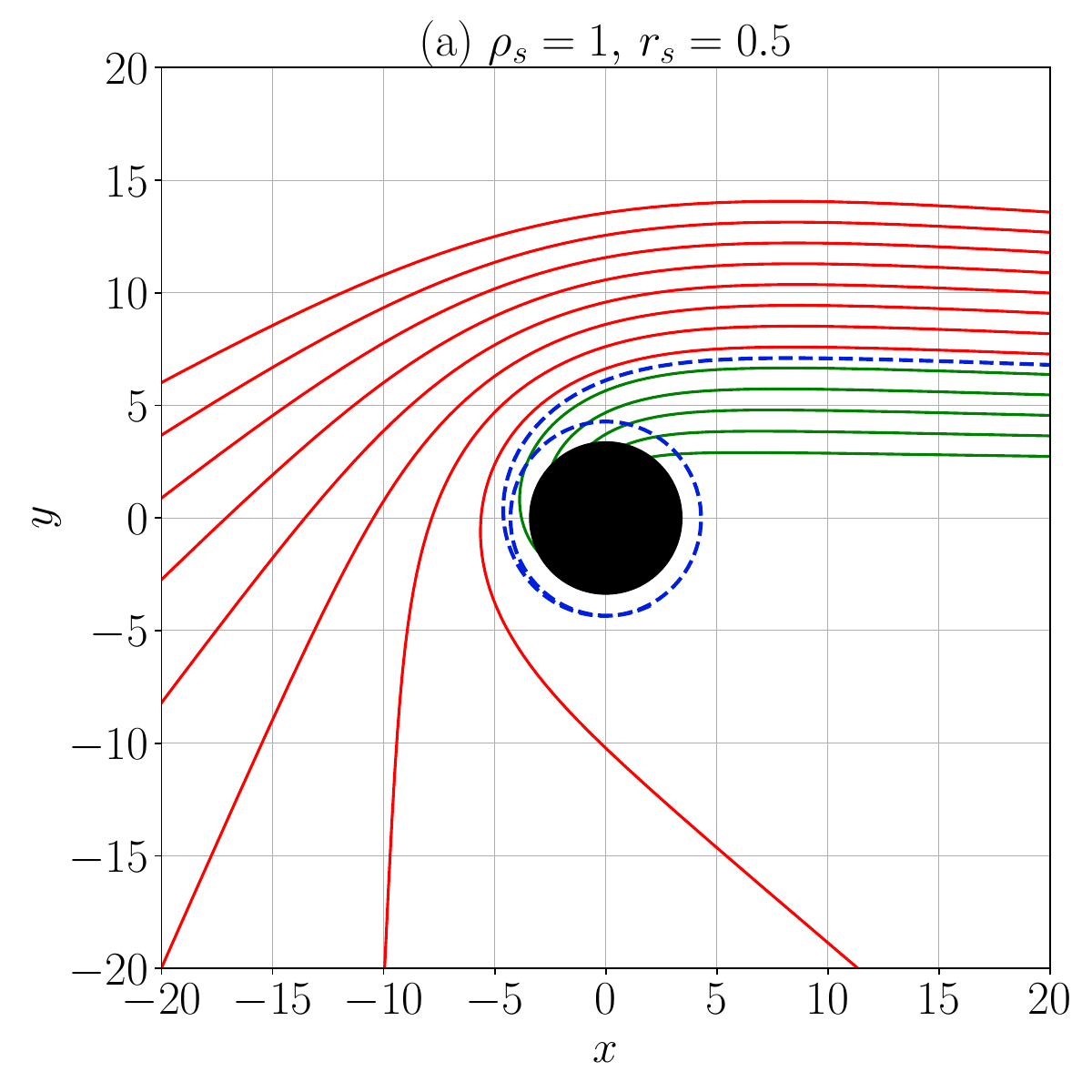}\hspace{-0.1cm}\includegraphics[scale=0.3]{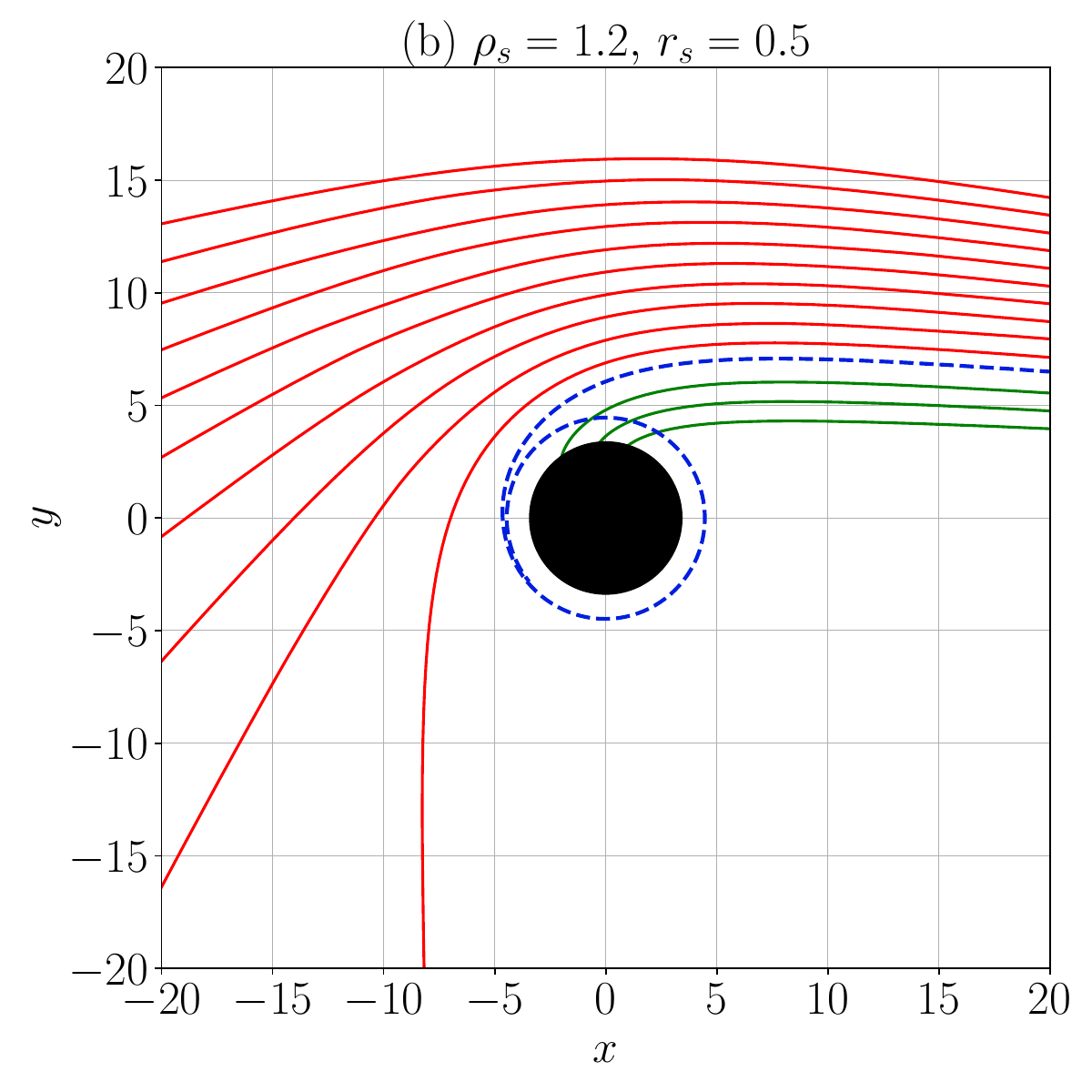} \hspace{-0.1cm}\includegraphics[scale=0.3]{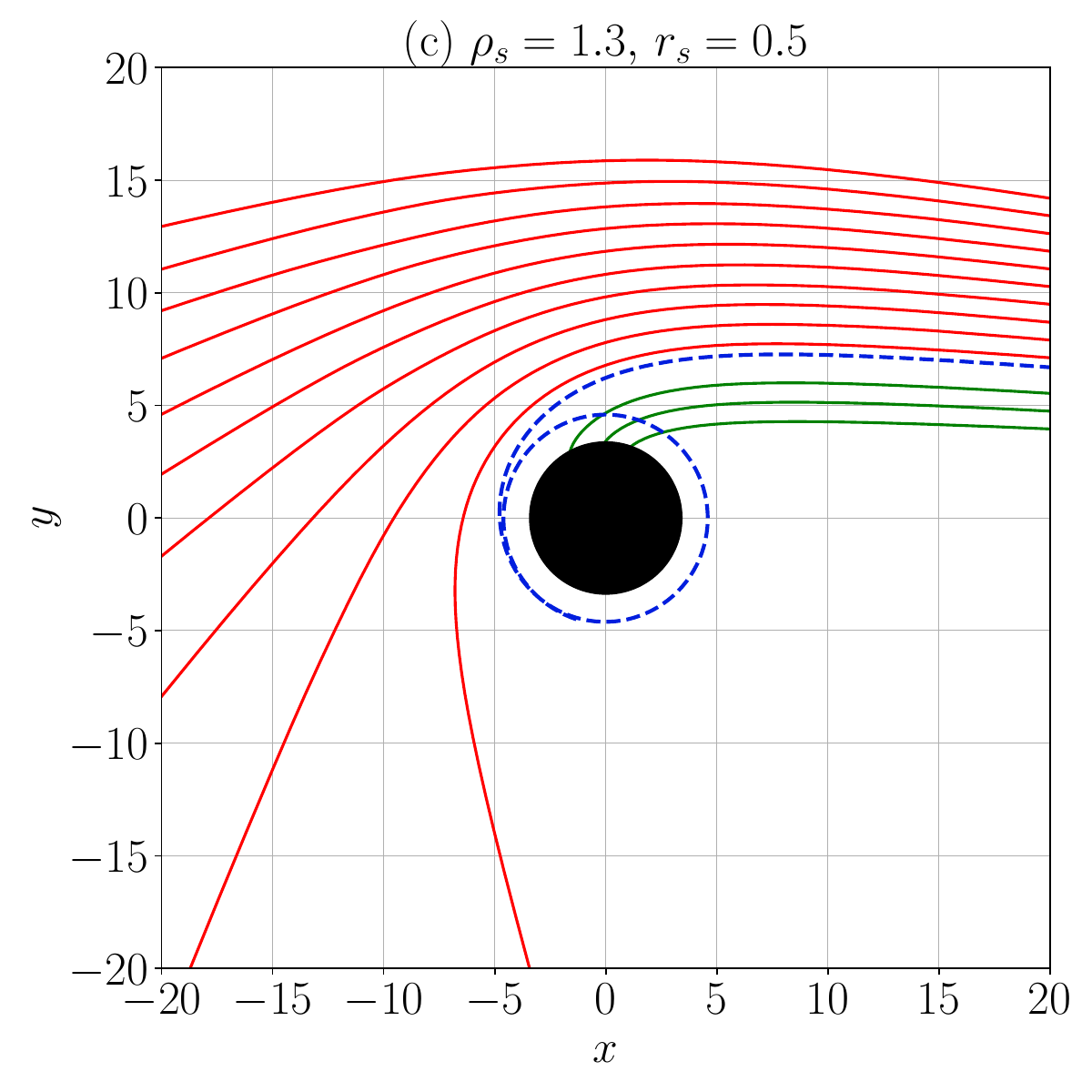}}
\caption{Null geodesics for different combination of $\rho_s$ and $r_s$.}
\label{null_geod}
\end{figure*}
\begin{figure*}[tbh]
\centerline{\includegraphics[scale=0.4]{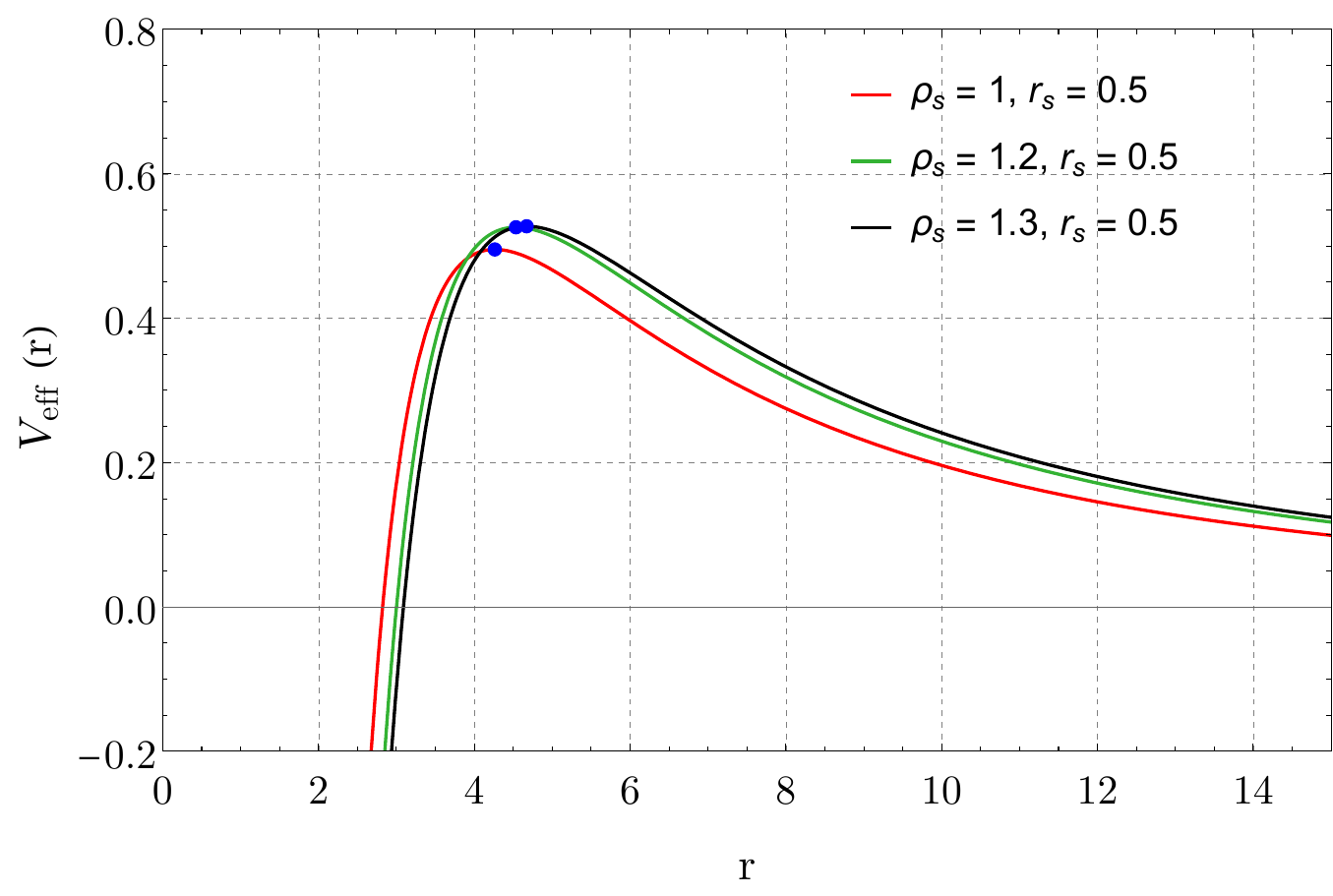}}
\caption{The plot of effective potential is shown for the different combinations of $\rho_s$ and $r_s$. The blue dots indicate the maximum of the potential and correspond to the blue-dashed circular orbits in Fig. \ref{null_geod}.}
\label{eff_pot_plot}
\end{figure*}
\begin{table*}[htb]
\centering
\begin{tabular}{||c|c|c||}
\hline
$\text{Photon sphere radius } (r_p)$  & $\text{Central core density} (\rho_s)$ & $\text{Impact parameter }(b)$      \\ \hline
$4.265$ & $1$      & $7.4708$ \\ \hline
$4.535$ & $1.2$      & $8.1993$ \\ \hline
$4.672$ & $1.32$   & $8.4631$ \\ \hline
\end{tabular}
\caption{Values of the radius of circular photon path for different values of the dark matter halo density parameter $\rho_s$ and impact parameter $b = L/E$.}
\label{tab_1}
\end{table*}
The null geodesics are obtained from the geodesic equations Eq. \eqref{geod_eqs_mod} by solving them simultaneously numerically and are shown in Fig. \ref{null_geod}. The circular photon orbits are represented by the blue dotted lines in the figure. The radius of the photon orbits corresponds to the peak of the effective potential depicted by the blue dots. The green trajectories indicate light rays with a lower impact parameter ($b = L/E$) that enter the BH. The red ones depict the paths of light beams with higher impact parameters. The circular orbit radius relates to a certain value of the impact parameter at which massless particles travel in a circular path. At any radius that is larger or smaller than the circular photon orbit radius, photons will escape or fall into the BH.

One can see clearly from Tab. \ref{tab_1}, that the radius of circular photon orbit increases with increasing dark matter halo core radius $\rho_s$. Moreover, the impact parameter corresponding to the circular orbits also increases with increasing values of $\rho_s$. This result shows that the presence of dark matter influences the circular geodesics of light rays around the BH. 

\subsection{Lyapunov Stability: Dynamical Systems Approach}
\begin{figure*}[tbh]
\centerline{\includegraphics[scale=0.5]{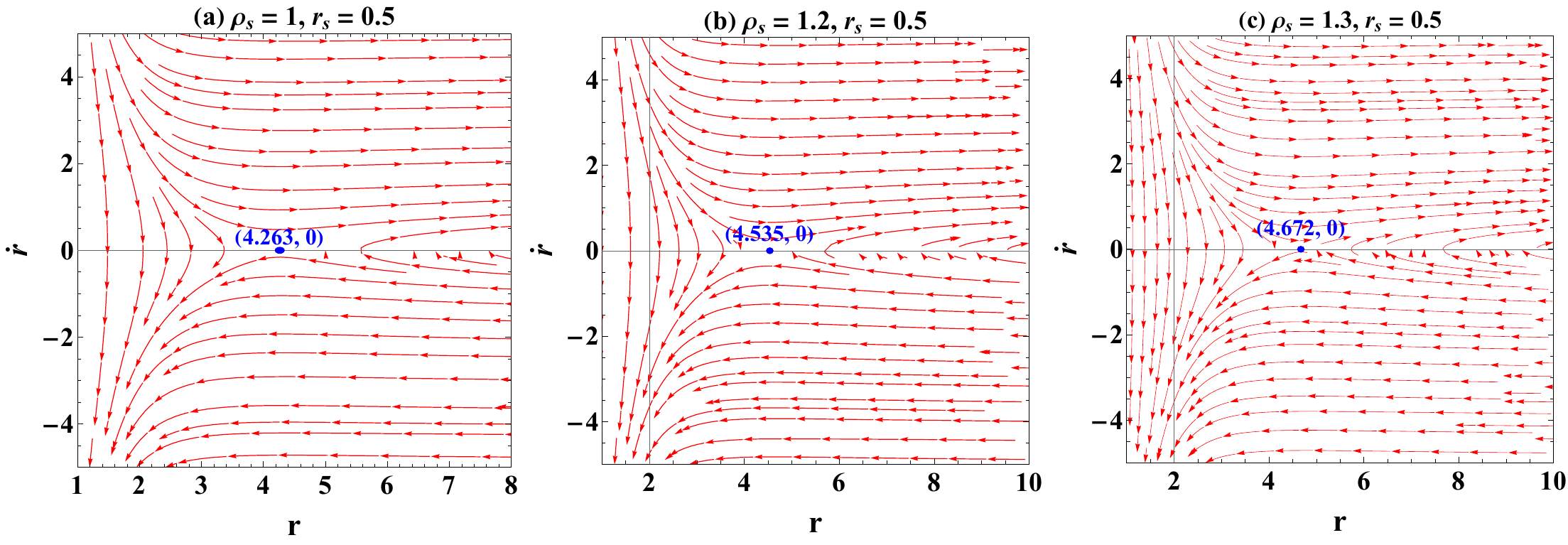}}
\caption{Phase portrait of $r$ vs $\dot{r}$ for the circular null geodesics for different values of $\rho_s$ and $r_s$.}
\label{phase_plot}
\end{figure*}

To analyze the stability of null circular geodesics using the Lyapunov method, we begin by constructing a dynamical system and examine its phase structure in the $(r, \dot{r})$ plane. For the case of circular geodesics, it is anticipated that $\dot{r} = 0$ which reduces the phase plane to the $(r, 0)$ plane. By observing the phase flow dynamics in the $(r, 0)$ plane, information about the critical point $(r_c, 0)$ can be obtained. To do that we differentiate Eq. \eqref{eff1} and then eliminate $\dot{r}$ to get
\begin{equation}
\ddot{r} = - \frac{d V_{eff}}{dr},
\label{rddot}
\end{equation}
We choose the coordinates $x_1 = \dot{r}$ and $x_2 = \dot{x_1}$, which gives the following set of differential equations
\begin{equation}
\begin{aligned}
x_1 &= \dot{r} \\
x_2 &= - \frac{d V_{eff}}{dr} = \frac{1}{3 r^4 \left(r_s+r\right)^3}\left[L^2 \left(-9 M \left(r_s+r\right)^3+3 r^4+9 r^3 r_s+3 r^2 r_s^2 \left(8 \pi  r_s^2 \rho _s+3\right)\right. \right. \\& \left. \left. \hspace{8cm}+12 \pi  r_s^6 \rho _s+r r_s^3 \left(32 \pi  r_s^2 \rho _s+3\right)\right)\right].
\end{aligned}
\label{dyn_sys}
\end{equation}
The Jacobian matrix $\mathcal{J}$ of the system \eqref{dyn_sys} is 
\begin{equation}
\mathcal{J} = \begin{pmatrix}
0 & 1 \\
- V_{eff}''(r) & 0
\end{pmatrix}
\label{jac}
\end{equation}
 where $''$ denotes the double differentiation with respect to $r$. The eigenvalue equation is 
 \begin{equation}
 \lambda^2 = - V_{eff} '' (r),
 \label{eig1}
\end{equation}
It can be seen that when $V_{eff}''(r) > 0$, then $\lambda^2 < 0$, which implies that the critical point represents a stable center point whereas when $V_{eff}''(r) < 0$, $\lambda^2 > 0$, which represents a saddle critical point. The phase portrait is shown in Fig. \ref{phase_plot} for three combinations of the dark matter halo core density $\rho_s$ and central core radius $r_s$.
The phase portrait shows the phase flow of the circular null geodesics for different combinations of $\rho_s$ and $r_s$. The circular geodesics corresponding to the photon sphere radius are anticipated to be unstable at the peak of the effective potential for all these combinations. The instability arises in the sense that even the tiniest perturbation to this orbit makes the photons lose their trajectory, ultimately plunging into the event horizon of the BH. This scenario is illustrated in Fig. \ref{phase_plot}, where the unstable circular orbits are depicted as saddle critical points. For \(\rho_s = 1\) , \(\rho_s = 1.2\) and \(\rho_s = 1.3\), the critical points occur at \(4.265,0)\), \((4.535, 0)\) and \((4.672,0)\) respectively in the $r - \dot{r}$ phase plane. If photons are at one of these saddle points, the tiniest perturbation can cause them to deviate from their circular path and can lead them either into the BH or away from it. This explains the trajectories diverging in the phase plane. This showcases the sensitivity of these photon orbits to infinitesimal perturbations. This instability arises as a consequence of the second derivative of the effective potential at the point corresponding to the circular photon orbit.

\subsection{Stability of Circular Geodesics: Lyapunov Exponent}
The Lyapunov exponent gives the average rate of divergence between two proximate geodesics within the phase plane. A positive value of the Lyapunov exponent represents the divergence of these geodesics, while a negative value represents their convergence. In the context of BH spacetimes, the presence of unstable circular geodesics marks the non-linear nature of GR, represented by a positive value of the Lyapunov exponent. This non-linearity indicates that the system is non-integrable, and the circular geodesics may exhibit chaotic behaviour. The Lyapunov exponent ($\tilde{\lambda}$) is directly related to the effective potential through the relation \cite{Kumara2024Jan}
\begin{equation}
\tilde{\lambda}^2 = -\frac{(V_{eff})''}{2\dot{t}^2},
\label{lyap}
\end{equation}
where ($^{\prime \prime}$) represents the second derivative with respect to $r$. 
The circular geodesics are unstable, stable and marginally stable for real, imaginary and zero values of $\tilde{\lambda}$. For our model, we find the Lyapunov exponent as
\begin{equation}
\begin{aligned}
\tilde{\lambda}^2 &= \frac{1}{2 E^2 r_c^5 \left(r_c+r_s\right)^4}\left[L^2 \left(\frac{2 M}{r_c}+\frac{4 \pi  r_s^3 \rho _s \left(2 r_c+r_s\right)}{3 \left(r_c+r_s\right){}^2}-1\right){}^2 \left(12 M \left(r_c+r_s\right){}^4+r_c \left(4 r_c^3 r_s \left(4 \pi  r_s^2 \rho _s-3\right) \right. \right. \right. \\& \left. \left. \left. \hspace{3cm}+6 r_c^2 r_s^2 \left(4 \pi  r_s^2 \rho _s-3\right)+4 r_c r_s^3 \left(4 \pi  r_s^2 \rho _s-3\right)-3 r_c^4+r_s^4 \left(4 \pi  r_s^2 \rho _s-3\right)\right)\right)\right]
\end{aligned}
\label{lyap2}
\end{equation} 
\begin{figure*}[tbh]
\centerline{\includegraphics[scale=0.4]{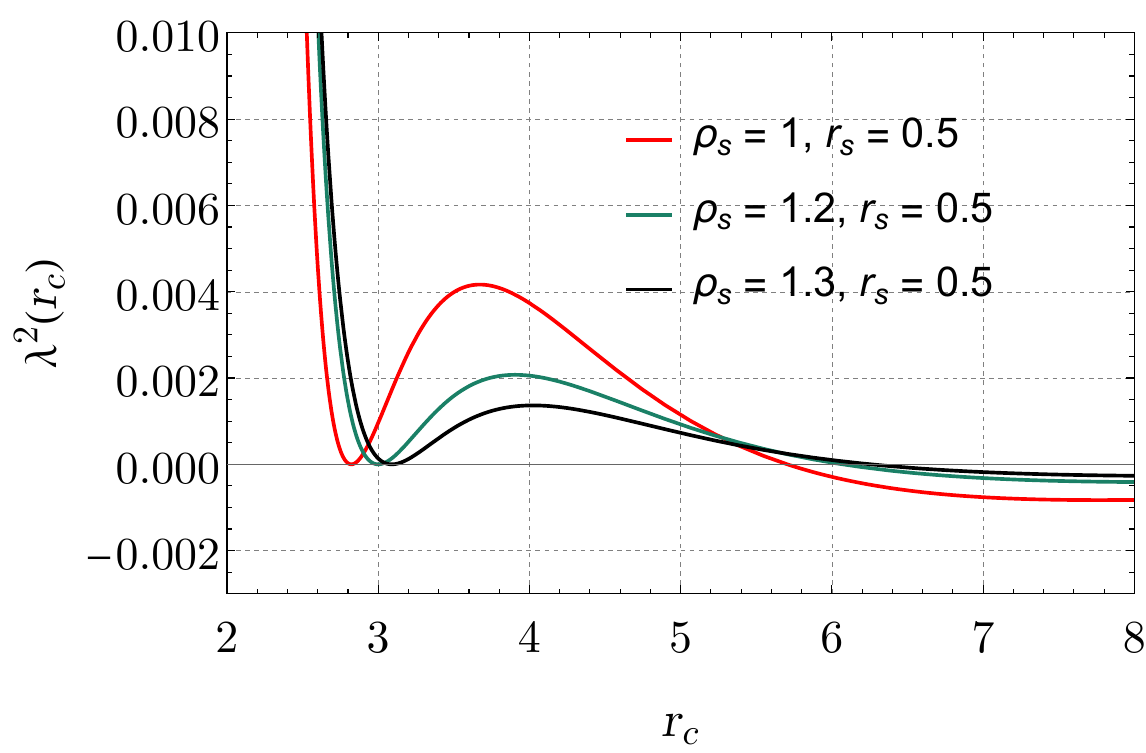}}
\caption{The plot of Lyapunov exponent is shown for the different combination of $\rho_s$ and $r_s$}
\label{lyap_plot}
\end{figure*}
The variation of the Lyapunov exponent squared $\tilde{\lambda}^2$ with respect to the radius of circular orbit $r_c$ is advocated by the condition $\dot{r} = 0$. From Fig. \ref{lyap_plot}, it is apparent that the circular photon orbits are unstable at smaller radii and become marginally stable at larger radii. From Fig. \ref{lyap_plot}, it can be clearly seen that for smaller photon sphere radius, the Lyapunov exponent is positive which explains the unstable nature of circular geodesics, which conforms with the results obtained from the backward ray tracing method and the phase portrait from Figs. \ref{null_geod} and \ref{phase_plot} respectively. Moreover, the circular geodesics becomes stable for the larger radius of circular orbits as represented by negative values of the Lyapunov exponent.

\begin{figure*}[tbh]
\centerline{\includegraphics[scale=0.5]{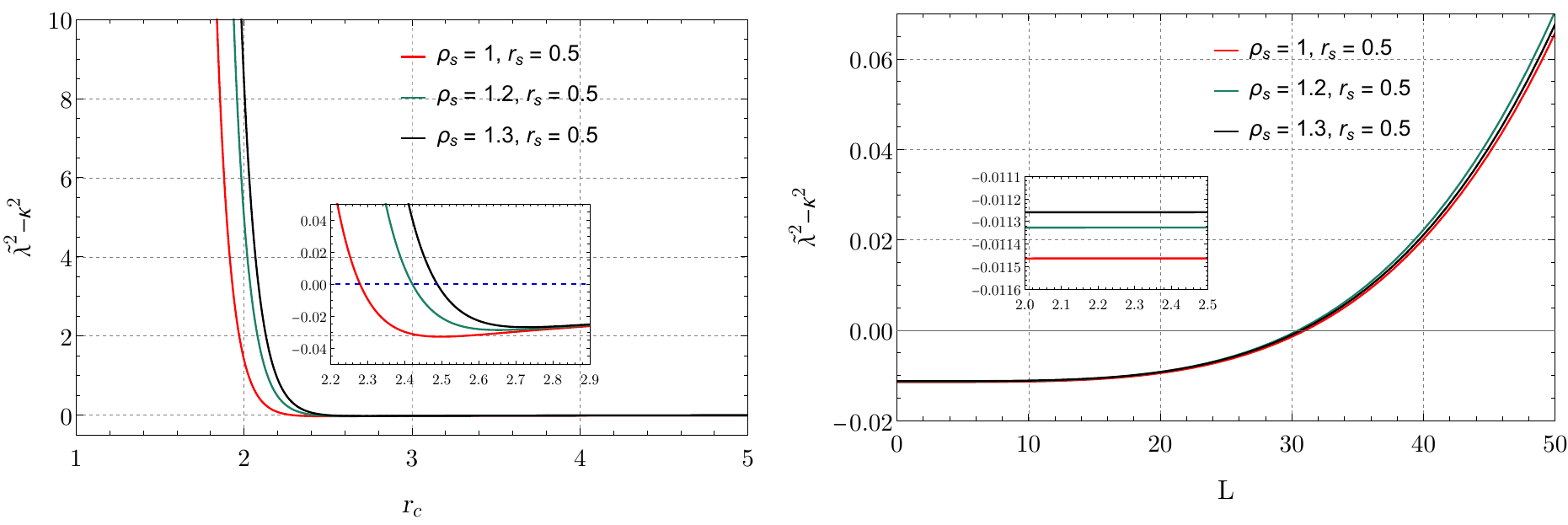}}
\caption{The plot of effective potential is shown for the different combination of $\rho_s$ and $r_s$}
\label{chaos_bound_fig}
\end{figure*}

This result can also be supported by the investigation of the chaos bound of the null circular geodesics. Chaos is a phenomenon observed in physical systems, representing unpredictable and random motion in deterministic and non-linear dynamical systems arising from the sensitivity to the initial conditions. From the study of chaotic systems, it was revealed that if chaos is mathematically represented as a function of time $C(t)$, it increases exponentially over time, i.e. \( C(t) \approx \exp(\tilde{\lambda} t) \), where \( \tilde{\lambda} \) is the Lyapunov exponent, which is the manifestation of the system's sensitivity on the initial conditions. Maldacena, Shenker and Stanford proposed a conjecture that there should be a universal upper bound for the Lyapunov exponent in thermodynamic systems \cite{Maldacena2016Aug}
\begin{equation}
\tilde{\lambda} \le \frac{\kappa}{\hbar},
\label{chaos_bound}
\end{equation}
where $\kappa$ is the surface gravity of the BH which is related to the temperature of the BH through $T=\kappa/2\pi$. Considering units where $\hbar = 1$, one may calculate the quantity $\tilde{\lambda}^2 - \kappa^2$. It becomes obvious that if $\tilde{\lambda}^2 - \kappa^2 < 0$, the chaos bound condition is satisfied whereas if  $\tilde{\lambda}^2 - \kappa^2 > 0$, the chaos bound is violated. In our model the evolution of $\tilde{\lambda}^2 - \kappa^2$ is shown with respect to the radius of photon orbits and angular momentum $L$ in Fig. \ref{chaos_bound_fig}. 
It is observed from the figure that, the chaos bound is violated for smaller photon orbit radii and satisfied for larger orbits. This implies that chaotic behaviour of the orbits may be witnessed near the vicinity of the BH. Moreover, it is also seen that chaos bound is obeyed for smaller values of angular momentum of the photons, and violated for larger values. This suggests that orbits of higher angular momentum photons may display chaotic behaviour. The violation of the chaos bound for smaller photon orbit radii or in the vicinity of the BH implies modification of the thermodynamic properties of the BH in dark matter environments. Since, it is clearly seen how the dark matter parameters affect the temperature of the BH in our analysis in section \ref{sec3}, it is possible that the presence of dark matter requires revisiting or extending the standard thermodynamics of BH based on Bekenstein-Hawking entropy. Also, the violation of chaos bound at smaller photon orbits hints at the possibility of new physics beyond GR in the regime of strong gravitational fields in the presence of dark matter.

\section{Conclusion}
\label{conc}
The composition of BH and dark matter can be an intriguing physical system, conceivably important from the viewpoint of BHT. In the initial part of the paper, we derive the functional form of the BH metric function by considering a Schwarzschild BH embedded within a Dehnen-type dark matter halo profile. We then examine the thermodynamic properties of this resulting system, focusing on the thermodynamic variables like Hawking temperature, specific heat, and free energy. The main results are as follows: \textbf{(a)} For a fixed value of the horizon radius within a physically allowed range, BHs surrounded by a dark matter halo with a lower central core density exhibit higher temperatures. As the core density parameter of the dark matter halo increases, the peak value of the temperature ceases to exist, indicating no phase transition for higher values of central core density.
\textbf{(b)} We found that owing to the influence of the dark matter halo, the BH is locally stable for smaller BHs and unstable for larger BHs after a phase transition. Thermodynamically stable BH states tend to disappear for higher values of the central core density parameter of the dark matter halo. This restricts the allowed values of the core density. \textbf{(c)} The horizon radii at which the phase transition occurs seem to increase with the increase in the central core density parameter. Additionally, with the influence of the dark matter halo parameters, we found that smaller BHs are globally thermodynamically stable and unstable for larger BHs. 
\textbf{(d)} We derived the effective potential of the BH system and the null geodesic equations resulting from our obtained BH solution. By numerically analyzing the null geodesics using the method of backward ray-tracing of light rays around the BH, we observe that the radius of the unstable circular orbit of light rays increases with increasing dark matter halo core density. The photon orbit radii appear as the critical points in the phase space, whose nature happens to be saddle points representing unstable photon orbits. \textbf{(e)} Finally, the chaotic nature of the circular photon orbits is studied using Lyapunov exponents and chaos bound. We see that for smaller photon orbit radius, the Lyapunov exponent takes positive values which explains the unstable nature of circular geodesics, which conforms with the findings obtained from the backward ray tracing method and the phase portrait mentioned in point \textbf{(d)}. \textbf{(f)} The chaos bound is violated for smaller photon orbit radii and satisfied for larger orbits. Thus, the chaotic behaviour of the orbits results in the vicinity of the BH. It is also found that the chaos bound is obeyed for smaller values of angular momentum of the photons, and violated for larger values. This suggests that orbits of higher angular momentum photons may display chaotic behaviour. 

The present work primarily deals with the thermodynamic stability of the dark matter-BH system and the nature of null geodesics around the BH. Also, we have not investigated the effect of charge and rotation of the BH which by the way can be intriguing to investigate and is left as a future scope. The introduction of charge and rotation in the current system will introduce a number of new parameters, that can open new possibilities with a more diverse physical system both in terms of thermodynamics and optical properties. Other perspectives can be carried out like quasinormal modes, shadows, thermodynamical phase structures and topology which are currently beyond the scope of this paper.  We intend to investigate the same in future. 

\section*{Acknowledgment}
The authors thank Hassan Hassanabadi for his insightful suggestions that helped us improve the paper significantly.

\bibliography{bibliography.bib}
\end{document}